\documentclass[notitlepage,aps,11pt,amsmath,amsfonts,amssymb,nofootinbib,superscriptaddress]{revtex4-1}
\usepackage{enumerate}
\usepackage{dsfont}
\usepackage{amsfonts}
\usepackage{amssymb}
\usepackage[latin1]{inputenc}
\usepackage[T1]{fontenc}
\usepackage{graphicx}
\usepackage{mathtools}
\usepackage{bbold}		

\usepackage[dvipsnames]{xcolor}

\usepackage[colorlinks=true]{hyperref}
\usepackage{amsthm}

\begin{document}
\title{Planck-scale-modified dispersion relations in homogeneous and isotropic spacetimes}

\author{Leonardo Barcaroli}
\email{leonardo.barcaroli@roma1.infn.it}
\affiliation{Dipartimento di Fisica, Universit\`a "La Sapienza''
and Sez. Roma1 INFN, P.le A. Moro 2, 00185 Roma, Italy}

\author{Lukas K. Brunkhorst}
\email{ lukas.brunkhorst@zarm.uni-bremen.de}
\affiliation{Center of Applied Space Technology and Microgravity (ZARM),
   University of Bremen, Am Fallturm, 28359 Bremen, Germany}

\author{Giulia Gubitosi}
\email{g.gubitosi@imperial.ac.uk}
\affiliation{Theoretical Physics, Blackett Laboratory, Imperial College, London SW7 2AZ, United Kingdom.}

\author{Niccol\'o Loret}
\email{niccolo.loret@roma1.infn.it}
\affiliation{Division of Theoretical Physics, Ru\dj er Bo\v{s}kovi\'{c} Institute, Bijeni\v{c}ka c.54, 10000 Zagreb, Croatia}

\author{Christian Pfeifer}
\email{christian.pfeifer@itp.uni-hannover.de}
\affiliation{Institute for Theoretical Physics, Universit\"at Hannover, Appelstrasse 2, 30167 Hannover, Germany}
\affiliation{Center of Applied Space Technology and Microgravity (ZARM), University of Bremen, Am Fallturm, 28359 Bremen, Germany}

\begin{abstract}
The covariant understanding of dispersion relations as level sets of Hamilton functions on phase space enables us to derive the most general dispersion relation compatible with homogeneous and isotropic spacetimes. We use this concept to present a Planck-scale deformation of the Hamiltonian of a particle in Friedman-Lema\^itre-Robertson-Walker (FLRW) geometry that is locally identical to the $\kappa$-Poincar\'e dispersion relation, in the same way as the dispersion relation of point particles in general relativity is locally identical to the one valid in special relativity. Studying the motion of particles subject to  such Hamiltonian we derive the redshift and lateshift as observable consequences of the Planck-scale deformed FLRW universe.
\end{abstract}

\maketitle

\section{Introduction}
The most promising scenarios for testing quantum gravitational effects are within astrophysical or cosmological settings \cite{AmelinoCamelia:2008qg, Mattingly:2005re}, described by spacetimes with non-vanishing curvature like the Friedman-Lema\^itre-Robertson-Walker (FLRW) spacetime. Thus, the issue of identifying a suitable framework for the phenomenological description of Planck-scale physics in curved spacetimes is very important in order to make contact with observations.

In particular, the possibility that the dispersion  relation of free relativistic particles is modified at the Planck scale is one of the main areas of research in quantum gravity phenomenology \cite{AmelinoCamelia:2008qg, Mattingly:2005re,Liberati:2013xla}. Most of the model building efforts so far focused on modifications of the special relativistic dispersion relation satisfied by particles moving on a flat spacetime. In this context, there have been several proposals, connected to different underlying assumptions concerning the fate of relativistic symmetries at the Planck scale. In Lorentz-violating (LIV) models, the deformed dispersion relation results in the introduction of a preferred frame, the one in which the dispersion relation takes the given form. Such LIV models can be straightforwardly extended to the case of a curved spacetime, just by introducing the appropriate redshift factors for energy and momentum \cite{Jacob:2008bw}. 
In other cases deformed dispersion relations do not introduce preferred reference frames, thanks to an appropriate modification of the laws of transformation between inertial observers, so that the deformed dispersion relation is invariant under deformed Lorentz transformations. Such models are generically known as 'Doubly (or Deformed) Special Relativity' (DSR) \cite{AmelinoCamelia:2000mn, AmelinoCamelia:2000ge, KowalskiGlikman:2001gp, Magueijo:2002am} models. It has been understood that these generically entail a curved  momentum space \cite{KowalskiGlikman:2002ft, KowalskiGlikman:2003we, AmelinoCamelia:2011bm, Gubitosi:2013rna, Arzano:2014jua}, which has (anti)-de Sitter geometry in order to be symmetric under the full set of modified Lorentz transformations. 
Within the DSR framework, a much studied class of models describes the deformed symmetries in terms of  Hopf algebras rather than Lie algebras \cite{Majid:1994cy, KowalskiGlikman:2001px}. In particular, the $\kappa$-Poincar\'e Hopf algebra \cite{Lukierski:1991pn, Lukierski:1992dt, Lukierski:1993wxa} has arguably received the highest attention from a phenomenological point of view. For this algebra, in the so-called bicrossproduct basis \cite{Majid:1994cy}, the dispersion relation for a free particle takes the form:
\begin{equation}
-\left(\frac{2}{\ell}\right)^2\sinh^2\left(\frac{\ell}{2} E\right)+e^{\ell E}|\vec p|^2=-m^2\,,\label{eq:kP}
\end{equation}
where $\ell$ is the quantum deformation parameter. The energy/momentum variables $E,p$ live on a manifold with de Sitter geometry \cite{Gubitosi:2013rna,KowalskiGlikman:2002ft,KowalskiGlikman:2003we,AmelinoCamelia:2011bm}, while spacetime is flat. So when studying the phenomenology of the $\kappa$-Poincar\'e algebra it is convenient to look at the momentum space as the base manifold, and at spacetime as the fiber of its cotangent bundle. Then it is possible to derive features such as the difference in travel time of free particles with different energies (lateshift) using methods that are completely analogous to the one used to compute the redshift of particles in general relativity \cite{Amelino-Camelia:2013uya}, just exchanging the role of spacetime and momentum space as base manifolds. 
Problems arise when attempting to introduce DSR effects in situations where spacetime is curved \footnote{When spacetime is curved (eventually deformed) Poincar\'e symmetries hold locally.}. In this case both of the spacetime and momentum space manifolds would be curved and their geometries intertwined \cite{Marciano:2010gq,Cianfrani:2014fia,AmelinoCamelia:2012it}. 
A geometrical treatment becomes challenging, and approaches such as the above-mentioned Hopf algebras have a limited applicability because they assume the existence of global symmetries. In fact, the only case that admits a Hopf-algebraic description is the one where spacetime has de Sitter geometry. The quantum-deformed de Sitter Hopf algebra is known as $q$-de Sitter algebra \cite{Ballesteros:2004eu, Ballesteros:2013fh, Barcaroli:2015eqe}. It is such that in the limit of flat spacetime it reduces to the $\kappa$-Poincar\'e Hopf algebra and momentum space has de Sitter geometry.

In previous work \cite{Barcaroli:2015xda} we proposed to use the framework of a Hamilton geometry of phase space to introduce modified dispersion relations within a generic curved spacetime in such a way that (possibly modified) relativistic symmetries are not spoiled.  We demonstrated how a dispersion relation can be understood covariantly as level set of a Hamilton function on point particle phase space, or, mathematically speaking, on the cotangent bundle of spacetime. 
We also showed that the Hamilton geometry  framework is suitable to describe the kinematics of free particles as well as the (eventually modified) symmetry properties of the dispersion relation. As an example, we  discussed the application of the Hamilton geometry framework to dispersion relations inspired from the $\kappa$-Poincar\'e  Hopf algebra and from the $q$-de Sitter Hopf algebra.

Of course, the ultimate scope of introducing the Hamilton geometry framework is to deal with more realistic scenarios involving less symmetric spacetimes. In particular, the Friedman-Lema\^itre-Robertson-Walker geometry, which models the universe we live in on cosmological scales, 
is homogeneous and isotropic, but lacks time-translation symmetry. In the approach of ref. \cite{Rosati:2015pga} an FLRW model was constructed starting from an algebraic setting by combining slices of maximally symmetric de Sitter spacetimes. The suggested procedure is however hardly generalizable to other kinds of spacetimes. In this respect the Hamilton geometry approach seems to be better suited to describe Planck-scale deformations of general spacetimes in  a way that is compatible with their symmetry properties.
In this article we leave maximally symmetric spacetimes as building blocks behind and deal with less symmetric spacetimes, focussing on the homogeneous and isotropic case.
The precise notion of symmetry which we introduced in our previous work allows us to derive the most general form of a dispersion relation which is compatible with this reduced set of symmetries.
As a special case, we exhibit a Planck-scale-modified dispersion relation which is compatible with homogeneity and isotropy of spacetime and which locally reduces to the  $\kappa$-Poincar\'e dispersion relation \footnote{Of course, when the Planck-scale deformation is removed, this dispersion relation reduces to the one characterizing free particles in FLRW spacetime.}. In the Hamilton geometry framework, this dispersion relation yields a spacetime manifold $M$ equipped with a Hamilton function $H$ on its cotangent bundle $T^*M$ which together we call Planck-scale-deformed FLRW phase space \footnote{We can not talk separately of spacetime and momentum space because their modified geometrical properties are intertwined.}. 

Thanks to this description of a Planck-scale-deformed homogeneous and isotropic phase space within the Hamilton geometry setting, it is possible to look for observable consequences of the deformation by studying the phase space worldlines of free particles. This allows in particular to make predictions for the expected redshift and for the travel times of particles with different energies.

We present our results throughout this article as follows. We begin by recalling our interpretation of dispersion relations as level sets of Hamilton functions on the cotangent bundle of spacetime and their symmetries in section \ref{sec:symm}. In section \ref{sec:homiso} we present our main mathematical result, namely the most general homogeneous and isotropic dispersion relation, which is displayed in equation \eqref{eq:generalH}. To gain intuition about what can be said in general about the motion of free particles satisfying homogeneous and isotropic dispersion relations we investigate the Hamilton equations of motion induced by \eqref{eq:generalH} in section \ref{sec:HEOM}. Finally, we specialize to a Planck-scale deformation of FLRW spacetime geometry in section \ref{sec:qFLRW}. We compute the observable energy redshift and energy-dependent time of travel in subsections \ref{ssec:redshift} and \ref{ssec:lateshift}. These effects provide direct access to a comparison between the model and astrophysical observations.

In the following, $x$ stand for spacetime coordinates and $p$ for the four-momentum coordinates. $\partial_a$ and $\bar\partial^a$ stand, respectively, for derivation with respect to $x^a$ and $p_a$. The speed of light, $c$, is set to one and we use metric signature $(-,+,+,+)$.

\section{Dispersion relations as Hamilton functions and their symmetries}\label{sec:symm}
Dispersion relations can be identified with level sets of Hamilton functions $H(x,p)$ on phase space, where we identify phase space with the cotangent bundle~$T^*M$ of the spacetime manifold~$M$. In \cite{Barcaroli:2015xda} we discussed this procedure in detail and showed how to derive the geometry of phase space solely from Hamilton functions. To understand general relativity in this way observe that the general relativistic dispersion relation for a particle in free-fall
\begin{align}
- E^2 + \vec p\,^2 = - m^2
\end{align}
is nothing but the representation of a level set of the cotangent bundle function
\begin{align}
H_{GR}(x,p) = g^{ab}(x)p_ap_a = - m^2
\end{align}
in a local inertial frame. The phase space geometry derived from $H_{GR}$ then is nothing but a flat momentum space geometry and the usual Lorentzian metric geometry on spacetime.

An element of the cotangent bundle is a $1$-form $P$ on spacetime and gets labeled with so-called manifold induced coordinates $T^*M\ni P = p_a dx^a = (x,p)$. A coordinate change on spacetime $x \rightarrow \tilde x(x)$ induces a change of coordinates on $T^*M$ 
\begin{align}
(x,p) \rightarrow (\tilde x(x), \tilde p_a(x,p)) = \bigg(\tilde x(x), p_q \frac{\partial x^q}{\partial \tilde x^a}\bigg)\,,
\end{align}
while a local basis change from $dx^a$ to $\omega^b = A^{-1 b}{}_{a}(x)\,dx^a$ in a cotangent space $T^*_xM$ of $M$ is a \emph{linear}, possibly $x$-dependent, transformation in the $p$ part of the manifold induced coordinates on $T^*M$ only:
\begin{align}\label{eq:localbasischange}
(x,p)_{dx} = (x,\tilde p_a)_\omega = (x, p_q A^q{}_a(x))_\omega\,,
\end{align}
where the subscripts denote the basis in which we coordinatized the cotangent bundle. The coordinate basis $dx$ is usually suppressed. We mention these transformations here in some detail since we will use them later in section~\ref{sec:qFLRW} to justify our conjecture for the specific form of a quantum deformed cosmological Hamilton function. 

Since we seek to derive the most general dispersion relation whose symmetries are compatible with a homogeneous and isotropic spacetime we recall the notion of symmetry in the framework of Hamilton geometry \cite{Barcaroli:2015xda}.
Symmetries of a metric manifold $(M,g)$ are diffeomorphisms of the manifold which leave the metric invariant. 
In the same spirit we say that a diffeomorphism $\Phi$ of $T^*M$, i.e. of phase space, is a symmetry if it leaves the Hamiltonian (i.e. the dispersion relation) invariant:
\begin{equation}\label{eq:SymmetryCondition}
 H(\Phi(x,p))=H(x,p)\,. 
\end{equation}
From the infinitesimal action of the diffeomorphism,
\begin{equation}\label{eq:diffeo}
\Phi(x,p)=(\tilde x(x,p),\tilde p(x,p))=(x^a+\epsilon \xi^a(x,p),p_a+\epsilon \bar \xi_a(x,p))+\mathcal{O}(\epsilon^2)\,,
\end{equation}
one finds the vector field generating the symmetry transformation by asking that the above definition of symmetry, eq. (\ref{eq:SymmetryCondition}), holds:
\begin{eqnarray}
H(\Phi(x,p))&=&H(x^a+\xi^a(x,p),p_a+\bar \xi_a(x,p))\nonumber\\
&=&H(x,p)+\epsilon(\xi^a(x,p)\partial_a H(x,p)+\bar \xi_a(x,p)\bar{\partial}^aH(x,p))+\mathcal{O}(\epsilon^2)\nonumber\\
&=&H(x,p)+\epsilon Z(H)(x,p)+\mathcal{O}(\epsilon^2)=H(x,p)\,.
\end{eqnarray}
The vector field $Z=\xi^a(x,p)\partial_a+\bar \xi_a(x,p)\bar{\partial}^a$ on $T^*M$  is then the generator of the diffeomorphism $\Phi$, which has to satisfy the following condition for $\Phi$ to be a symmetry:
\begin{equation}\label{eq:symmcon}
Z(H)=0\,.
\end{equation}

In our previous work we identified several distinguished classes of symmetries. Here we are interested in manifold induced symmetries, which are symmetries of phase space induced by a diffeomorphism of the spacetime manifold.

A diffeomorphism of the base manifold $M$ can be represented infinitesimally by vector fields $X=\xi^a(x)\partial_a$ on $M$. It acts as a change of local coordinates $(x^a)\rightarrow(x^a+\xi^a)$. Such a local change of coordinates on $M$ induces a change of coordinates on $T^*M$ via $(x^a,p_a)\rightarrow(x^a+\xi^a,p_a-p_q\partial_a\xi^q)$. Thus a diffeomorphism on $M$ generated by the vector field $X$ induces a diffeomorphism on $T^*M$ generated by the vector field $X^C=\xi^a\partial_a-p_q\partial_a\xi^q\bar{\partial}^a$, the complete lift of $X$ from $M$ to $T^*M$. A symmetry of the Hamiltonian 
\begin{equation}\label{eq:mfsym}
	X^C(H)=0\,,
\end{equation}
which is determined by such a vector field $X$ on $M$ is called manifold induced symmetry.

As a remark, recall that manifold induced symmetries always lead to a conserved phase space functions \cite{Barcaroli:2015xda}
\begin{equation}\label{eq:conserved}
X^C(H)=0 \Leftrightarrow \{\xi^a(x)p_a,H\}=0\,,
\end{equation}
where $\{\cdot,\cdot\}$ are the standard Poisson brackets: $\{A,B\}\equiv \frac{\partial A}{\partial x^a}\frac{\partial B}{\partial p_a}-\frac{\partial B}{\partial x^a}\frac{\partial A}{\partial p_a}$.

We will now consider symmetry generating vector fields which generate homogeneous and isotropic spacetime manifolds and derive the most general  dispersion relation compatible with such symmetries.

\section{Homogeneous and Isotropic Hamiltonians}\label{sec:homiso}
For a homogeneous and isotropic spacetime we demand invariance under the diffeomorphisms induced by the following six vector fields encoding spatial rotations and spatial translations. In spherical spacetime coordinates $(t,r,\theta,\phi)$ and defining $\chi=\sqrt{1-kr^2}$, where $k$ is the spatial curvature, they take the form \cite{garecki}:
\begin{align}\label{eq:KVF1}
	X_1 &= \chi \sin\theta \cos\phi \,\partial_r + \frac{\chi}{r} \cos\theta \cos\phi\, \partial_\theta - \frac{\chi}{r}\frac{\sin\phi}{\sin\theta}\,\partial_\phi \nonumber\\
    X_2 &= \chi \sin\theta \sin\phi\, \partial_r + \frac{\chi}{r} \cos\theta \sin\phi \,\partial_\theta + \frac{\chi}{r}\frac{\cos\phi}{\sin\theta}\,\partial_\phi\\
    X_3 &= \chi \cos\theta \,\partial_r - \frac{\chi}{r}\sin\theta \,\partial_\theta \nonumber
\end{align}
for translations, and:
\begin{align}\label{eq:KVF2}
    X_4 &= \sin\phi \,\partial_\theta + \cot\theta \cos\phi \,\partial_\phi\nonumber\\
    X_5 &= -\cos\phi \,\partial_\theta + \cot\theta \sin\phi \,\partial_\phi\\
    X_6 &= \partial_\phi\,\nonumber
\end{align}
for rotations.

In order to find the most general homogeneous and isotropic Hamiltonian on phase space we need to derive the complete lifts of these vector fields to the cotangent bundle and demand that the Hamiltonian does not change along the flow of these vector fields, see \eqref{eq:mfsym}. The lifts $X^C_I, I=1,...,6$ of the above vector fields are quite long expressions and can be found in the appendix \ref{app:CL}.

There are two ways for evaluating the symmetry conditions $X^C_I(H)=0$ for $I=1, ..., 6$. One can solve the partial differential equations by  introducing the following set of new coordinates:
\begin{align}
	(\hat t, \hat r,\hat \theta, \hat \phi, \hat p_t, \hat p_r, \hat p_\theta, w)\,,
\end{align}
defined through
\begin{align}
	\hat t = t,\ \hat r = r,& \hat \theta = \theta,\ \hat \phi = \phi,\ \hat p_t = p_t,\ \hat p_r = p_r,\ \hat p_\theta = p_\theta,\\
    \label{eq:w2def}
	w^2 &= (1 - kr^2)p_r^2 + \frac{1}{r^2}p_\theta^2 + \frac{1}{r^2\sin\theta^2}p_\phi^2\,,
\end{align}
which is done in detail in appendix \ref{app:symcon}. Alternatively, the problem of solving the differential equations can be turned into an algebraic problem by taking an extrinsic point of view and seeing the $3$-dimensional homogeneous and isotropic spatial part of the geometry embedded into a $4$-dimensional ambient manifold. This is discussed in appendix \ref{sec:App-ExtrView}. 

Both the procedures yield that the most general Hamilton function on phase space which is spatially homogeneous and isotropic has the form
\begin{align}\label{eq:generalH}
	H(x,p) = H(t, p_t, w(r, \theta, p_r, p_\theta, p_\phi) )\,.
\end{align}
Thus we find that the dispersion relation is not allowed to depend arbitrarily on the momenta, but only on the energy component $p_t$ of the four-momentum and on the specific combination of spatial momenta encoded by $w$.

Some examples of homogeneous and isotropic Hamilton functions whose levels sets are then homogeneous and isotropic dispersion relations are:
\begin{itemize}
	\item The FLRW Hamiltonian:
	\begin{align}
			H_{FLRW} = - p_t^2 + a(t)^{-2} w^2\,,
    \end{align} 
where $a(t)$ is the scale factor appearing in the FLRW metric $$ds^2=-dt^2+a(t)^2 \left[ \frac{dr^2}{1-k r^2}+r^2\left(d\theta^2+\sin^2(\theta) d\phi^2\right)\right].$$

	\item A general first-order polynomial deformation of FLRW geometry:
	\begin{align}
			H_{dFLRW} =  - p_t^2 + a(t)^{-2} w^2 + \ell \big( b(t) p_t w^2 + c(t) p_t^3 \big) + \mathcal{O}(\ell^2)\,,\label{eq:dFLRW}
	\end{align}
where $a(t)$ is again the FLRW scale factor and $b(t), c(t)$ are generic functions of time. The parameter $\ell$ is the deformation parameter, with dimensions of length.     
   
	\item Our conjecture for a $\kappa$-Poincar\'{e}-inspired qFLRW dispersion relation, which will be motivated and studied in detail in section \ref{sec:qFLRW}:
	\begin{align}
		H_{qFLRW} = -\frac{4}{\ell^2}\sinh^2\bigg(\frac{\ell}{2} p_t\bigg) + a(t)^{-2} e^{\ell p_t}w^2\,,
	\end{align}
	whose first order expansion in $\ell$ reads:
	\begin{align}
		H_{qFLRW} = - p_t^2 + a(t)^{-2} w^2 + \ell a(t)^{-2} p_t w^2+ \mathcal{O}(\ell^2)\,.
	\end{align}
  So at the first order this corresponds to setting $b(t)=a(t)^{-2}$ and $c(t)=0$ in the general expression \eqref{eq:dFLRW}.
\end{itemize}

In section \ref{sec:qFLRW} we will study the qFLRW dispersion relation in more detail and derive its physical predictions. Before doing so we discuss the motion of particles subject to a general homogeneous and isotropic dispersion relation.

\section{The Hamilton equations of Motion}\label{sec:HEOM}
Homogeneous and isotropic Hamilton functions still possess a high degree of symmetry. Each generator of symmetry yields a constant of motion which enables us to simplify the Hamilton equations of motion significantly.

Expanding the equations of motion  $\dot p_a + \partial_a H = 0$ and $\dot x^a - \bar{\partial}^aH=0$ yields

\begin{minipage}{0.49999\textwidth} 
\begin{align}
		\dot p_t&=- \partial_t H\,,  \label{eq:ham1}\\
		\dot p_r&= \partial_wH \frac{1}{w}\bigg(kr p_r^2 + \frac{1}{r^3} p_\theta^2 + \frac{1}{r^3\sin\theta^2}  p_\phi^2 \bigg)\,,	\\
		\dot p_\theta &= \partial_wH \frac{1}{w} \frac{\cos\theta}{\sin\theta^3}p_\phi^2\,,\\
		\dot p_\phi&=0\,,
	\end{align}
\end{minipage}
	\hfill
\begin{minipage}{0.4\textwidth}
	\begin{align}
		\dot t &= \dot x^t = \bar\partial_t H\,,\\
		\dot r &= \dot x^r = \partial_wH \frac{1}{w} (1-kr^2)p_r\,,\\
		\dot \theta &= \dot x^\theta = \partial_wH \frac{1}{w} \frac{1}{r^2}p_\theta\,,\label{eq:dottheta}\\
		\dot \phi &= \dot x^\phi = \partial_wH \frac{1}{w} \frac{1}{r^2 \sin\theta^2} p_\phi\,\label{eq:dotphi}
        .
	\end{align}
\end{minipage}\\

The constants of motion \eqref{eq:conserved} can be obtained  in terms of the symmetry vector fields $X_I$ \eqref{eq:KVF1}-\eqref{eq:KVF2} via application of $X_I$ to a canonical $1$-form $P=p_a dx^a$. The conserved quantities associated to the generators of spatial rotations $X_4$ to $X_6$ are:
\begin{align}
	L_1 \equiv X_4(P) &= \sin\phi\, p_\theta + \cot\theta \cos\phi \,p_\phi\,,\nonumber\\
	L_2 \equiv X_5(P) &= -\cos\phi \,p_\theta + \cot\theta \sin\phi \,p_\phi\,,\\
	L_3 \equiv X_6(P) &= p_\phi\,,\nonumber
\end{align}
while the conserved quantities associated to the generators of spatial translations $X_1$ to $X_3$ are: 
\begin{align}
K_1 \equiv X_1(P) &= \chi \sin\theta \cos\phi p_r + \frac{\chi}{r}\cos\theta\cos\phi p_\theta - \frac{\chi}{r}\frac{\sin\phi}{\sin\theta}p_\phi \,,\nonumber\\
K_2 \equiv X_2(P) &= \chi \sin\theta \sin\phi p_r + \frac{\chi}{r}\cos\theta\sin\phi p_\theta + \frac{\chi}{r}\frac{\cos\phi}{\sin\theta}p_\phi \,,\\
K_3 \equiv X_3(P) &= \chi \cos\theta p_r - \frac{\chi}{r} \sin\theta p_\theta\,.\nonumber
\end{align}
Thanks to the symmetry properties of the system, in order to study geodesic motion it suffices to focus on radial geodesics along a given direction. One can then obtain the geodesics in all the other directions by applying a symmetry diffeomorphism, i.e. a  rotation or a translation.
Thus, without loss of generality we choose $\theta$ and $\phi$ to be constant and equal to 
$\theta=\frac{\pi}{2}$ and $\phi = 0$, which further simplifies the equations of motion to\\
\begin{minipage}{0.4\textwidth} 
	\begin{align}
	\dot p_t&=- \partial_t H\,,  \\
	\dot p_r&= \partial_wH \frac{1}{w}kr p_r^2 \,,\label{eq:dotpr}	\\
	\dot p_\theta &= 0\,,\\
	\dot p_\phi&=0\,,
	\end{align}
\end{minipage}
\hfill
\begin{minipage}{0.4\textwidth}
	\begin{align}
	\dot t &= \bar \partial_t H\,,\\
	\dot r & = \partial_w H \frac{1}{w} \chi^2 p_r\,,\\
	\dot \theta & = 0 
    \,,\\
	\dot \phi &= 0 
    \,.
	\end{align}
\end{minipage}\\

\noindent The two last equations imply the further constraints $p_\theta=p_\phi=0$ (compare with eqs. \eqref{eq:dottheta} and \eqref{eq:dotphi}).
Using this choice for $\theta$ and $\phi$, all the constants of motion vanish, except for $K_1$:
\begin{align}\label{eq:wcons}
    K_1 = \chi \ p_r\,.  
\end{align}
$K_1$ turns out to be related to the coordinate $w$ defined in \eqref{eq:w2def}, since when $\theta=\frac{\pi}{2}$ and $\phi=0$ this  becomes constant itself and satisfies $w^2(\theta=\frac{\pi}{2},\phi=0)= \chi^2 p_r^2=K_1^2$.

We reduced the dynamical system to have just four non-trivial equations, the ones for  $p_t, t$ and $p_r, r$. The last two of them can be further simplified  by observing that from \eqref{eq:wcons} it follows that:
\begin{align}\label{eq:prr}
	p_r = \frac{K_1}{\chi}=\frac{K_1}{\sqrt{1-k r^2}}\,.
\end{align}
Substituting this into the equation for $\dot r$ one gets:
\begin{equation}
\dot r =  \partial_w H \frac{1}{w} \chi  K_1\,.\label{eq:dotr1}
\end{equation}
One can of course verify that \eqref{eq:prr} and \eqref{eq:dotr1} are compatible with \eqref{eq:dotpr}.

Summarising, the dynamics of a  point particle with a homogeneous and isotropic Hamiltonian is described by the following set of equations for the phase space coordinates:

\begin{minipage}{0.4\textwidth} 
	\begin{align}
	\dot p_t&=- \partial_t H\,,\label{eq:dotpt}  \\
	p_r&=\frac{K_1}{\chi} \label{eq:pr} \,,	\\
	p_\theta &= 0\,,\\
	p_\phi&=0\,,
	\end{align}
\end{minipage}
\hfill
\begin{minipage}{0.4\textwidth}
	\begin{align}
	\dot t &= \bar \partial_t H\, \label{eq:dott}, \\
	\dot r & = \partial_wH \frac{1}{w} \chi K_1\, \label{eq:dotr},\\
	\theta & = \frac{\pi}{2}\,,\\
	\phi &= 0\,,
	\end{align}
\end{minipage}\\

\noindent where $K_1^2=w^2$ is a constant. For further analysis we need to specify the Hamiltonian we are interested in.

\section{The qFLRW dispersion relation}\label{sec:qFLRW}
In this last section we are going to study in detail one specific form of isotropic and homogeneous Hamiltonian. Our main reason for being interested in the Hamilton geometry framework is that it allows us to describe Planck-scale-modified dispersion relations as due to a deformed geometry of the phase space of relativistic particles, in a way that is compatible with the (deformed) symmetries of the dispersion relation. For this reason it makes sense to seek a generalisation of the $\kappa$- Poincar\'e Hamiltonian, eq. \eqref{eq:kP}, to a non-flat spacetime case.

As already briefly mentioned at the end of section \ref{sec:homiso}, a homogeneous and isotropic Hamiltonian that locally reduces to the $\kappa$-Poincar\'{e} one takes the form:
\begin{align}\label{eq:qFLRW}
	H_{qFLRW}(x,p) = -\frac{4}{\ell^2}\sinh^2\bigg(\frac{\ell}{2} p_t\bigg) + a(t)^{-2} e^{\ell p_t}w^2\,,
\end{align}
with Hamilton metric
\begin{align}
	g^{Hab} \equiv \frac{1}{2}\bar{\partial}^a \bar{\partial}^b H = 
	&\left(\begin{array}{cccc}
	-\cosh(\ell p_t) + \ell^2 e^{\ell p_t} \frac{w^2}{a^2} & \frac{\ell}{a^2} (1 - k r^2 ) e^{\ell p_t} p_r & \frac{\ell}{a^2} \frac{e^{\ell p_t}}{r^2} p_\theta & \frac{\ell}{a^2} \frac{e^{\ell p_t}}{r^2\sin\theta^2} p_\phi\\
	\frac{\ell}{a^2} (1 - k r^2 ) e^{\ell p_t} p_r &\frac{(1 - k r^2 )}{a^2} e^{\ell p_t}& 0  & 0\\ 
	\frac{\ell}{a^2} \frac{e^{\ell p_t}}{r^2} p_\theta & 0 & \frac{e^{\ell p_t}}{a^2 r^2}  & 0\\ 
	\frac{\ell}{a^2} \frac{e^{\ell p_t}}{r^2\sin\theta^2} p_\phi & 0 & 0  & \frac{e^{\ell p_t}}{a^2 r^2\sin\theta^2}\\ 
	\end{array}\right)\,.
\end{align}
In order to clarify in which sense this Hamiltonian is related to the $\kappa$-Poincar\'e one, we observe that at every point $x$ of spacetime there exists a local and linear transformation of the spatial momenta $p_{\alpha} \rightarrow \hat p_\alpha(p) = R^\beta{}_{\alpha}(r,\theta)p_\beta$, such that the coordinate $w$ takes the following form\footnote{(Greek indices run over spatial coordinates.)}:
\begin{align}
w^2 \rightarrow w^2 = \delta^{\alpha\beta}\hat p_\alpha \hat p_\beta\,.
\end{align}
This is true simply by the existence of local orthonormal frames for the spatial part of the FLRW metric and local frame transformations on $M$ are linear transformations on $T^*M$ as discussed around equation~\eqref{eq:localbasischange}. Thus, locally, at every $x\in M$ we find frame labels $(x,\tilde p)_\omega$ of phase space which are connected to the manifold-induced coordinates $(x,p)$ via a \emph{linear} transformation of the momenta:
\begin{align}
	p_\alpha \rightarrow \tilde p_\alpha =  a(t) \hat p_\alpha = a(t) R^\beta{}_\alpha(r,\theta) p_\beta \,.
\end{align}
In terms of these new momenta the Hamiltonian \eqref{eq:qFLRW} becomes:
\begin{align}\label{eq:qFLRWkP}
	H_{qFLRW}(x,\tilde p) = -\frac{4}{\ell^2}\sinh^2\bigg(\frac{\ell}{2}  \tilde p_t\bigg) +  e^{\ell \tilde p_t} \delta^{\alpha\beta}\tilde p_\alpha \tilde p_\beta\,,
\end{align}
which is the form of the Hamiltonian associated to the $\kappa$-Poincar\'e model, see eq. \eqref{eq:kP}.

The important observation is that this can be achieved by a \emph{linear} transformation in the momenta, which is analogous to the existence of local orthogonal inertial frames in general relativity, where at each point $x$ on spacetime there exists a local and linear transformation of the momenta such that the Hamiltonian of a Lorentzian spacetime $H_g(x,p)=g^{ab}(x)p_ap_b$ becomes $H_g(x,\tilde p) = \eta^{ab}\tilde p_a \tilde p_b$ \footnote{These linear transformations of the momenta are compatible with the Hamilton geometry of phase space which is derived from the Hamilton function since they are linear transformations, see \cite{Barcaroli:2015xda}.}. 

After having motivated our choice of qFLRW Hamilton function we can now proceed to discuss its observable physical consequences.

\subsection{Particle motion}
We already discussed the Hamilton equations of motion $\dot p_a + \partial_a H = 0$ and $\dot x^a - \bar{\partial}^aH=0$ for a general homogeneous and isotropic Hamiltonian in section \ref{sec:HEOM}. There we reduced the equations so that only the ones for four phase space variables are non-trivial. Of these, $p_r$ is given as a function of $r$ (eq. \eqref{eq:pr}), while $t$, $p_t$ and $r$ satisfy respectively the differential equations \eqref{eq:dott}, \eqref{eq:dotpt} and \eqref{eq:dotr}. When applied to the $qFLRW$ Hamiltonian \eqref{eq:qFLRW} and introducing $A(t) = a^{-2}(t)$ these equations read:
\begin{align}
\dot r &=  2 w A e^{\ell p_t} \chi\\
\dot p_t &= - A' e^{\ell p_t} w^2   \,,\\
\dot t &=  - \frac{2}{\ell}\sinh(\ell p_t) + \ell A e^{\ell p_t} w^2 \,,
\end{align}
where $w = \chi p_r = K_1$ is a constant of motion, see eq. \eqref{eq:wcons}, and the $'$ denotes derivative with respect to the coordinate time $t$.

We can avoid solving the differential equation for $p_t$ explicitly by using the fact that the Hamiltonian $H_{qFLRW}$ is conserved along the solutions of the equations of motion. This allows us to write $p_t(\tau)$ as a function of coordinate time $p_t(t)$. In fact, the Hamiltonian encodes the mass-shell condition for the particle:
\begin{align}
	H_{qFLRW}=-\frac{4}{\ell^2}\sinh^2(\frac{\ell}{2} p_t) + A e^{\ell p_t}w^2 =- m^2\,. 
\end{align}
By solving this with respect to $p_t$ one finds: 
\begin{align}\label{eq:pt(t)}
	p_t(t) = \frac{1}{\ell}\ln\bigg(\frac{2 + \ell^2 m^2 \pm \ell \sqrt{4 m^2 + \ell^2 m^4 + 4 w^2 A}}{2(1-\ell^2 w^2 A)}\bigg)\rightarrow \frac{1}{\ell}\ln\bigg(\frac{2 + \ell^2 m^2 - \ell \sqrt{4 m^2 + \ell^2 m^4 + 4 w^2 A}}{2(1-\ell^2 w^2 A)}\bigg)\,,
\end{align}
where the plus sign was chosen in order to recover $p_t = -m$ for a particle at rest in a flat spacetime in the $\ell=0$ limit, as it must be for our signature convention $(-,+,+,+)$. Note that $p_t(t)$ also depends on $p_r$ and $r$ via $w$.

We can also combine the differential equations for $r$ and $t$ in order to get the evolution of the radial coordinate with respect to the coordinate time $t$:
\begin{align}
r'(t)&\equiv\frac{d r}{d t} =\frac{\dot r}{\dot t} =  \frac{2 A e^{\ell p_t} \chi w}{\ell A e^{\ell p_t} w^2 - \frac{2}{\ell}\sinh(\ell p_t)} 
 = \frac{A \chi w ( - 2 - \ell^2 m^2+ \ell \sqrt{4 m^2 + \ell^2 m^4 + 4 w^2 A})  }{(\ell^2 w^2 A - 1)\sqrt{4 m^2 + \ell^2 m^4 + 4 w^2 A}}\,,\label{eq:rprime}
\end{align}
where in the last equality we used eq. \eqref{eq:pt(t)}.These equations for $p(t)$ and $r'(t)$, together with eq. \eqref{eq:prr} for $p_r$, fully characterize the motion of a particle and can be used to derive the relevant observables, as we show in the following subsection.

\subsection{Observable effects}
In order to characterize the physical predictions of the $H_{qFLRW}$ Hamiltonian it is useful to study the  behaviour of massless particles. In this case, both of the equations \eqref{eq:pt(t)} and \eqref{eq:rprime}  simplify significantly:
\begin{align}\label{eq:pt(t)massless}
	p_t(t) &= -\frac{1}{\ell}\ln\left(1+\ell w a^{-1}\right)\,,\\
    r'(t)&= \frac{\chi a^{-1}}{\ell w a^{-1} +1}\,.\label{eq:rprimemassless}
\end{align}
The evolution of $p_r$ can be straightforwardly derived from the one of $r$ using eq. \eqref{eq:prr}.

In the following we will study the evolution of the energy and radial coordinate implied by the above equations to show the emergence of two peculiar physical effects. One is a correction to the usual energy redshift, which turns out to depend on the initial energy of the particle besides its time of travel. The other effect is dubbed 'lateshift' \cite{Amelino-Camelia:2013uya,Barcaroli:2015eqe}, and consists in the fact that the time of travel of a particle depends on its energy, so that two particles emitted by a source at the same time but with different energies will arrive at a far away detector at different times. Both of these effects are already known in the context of studies of the phenomenology of particles whose symmetries are described by the $\kappa$-Poincar\'e Hopf algebra \cite{Amelino-Camelia:2013uya} or the $q$-de Sitter Hopf algebra \cite{Barcaroli:2015eqe}. In this phenomenological context, these Hopf algebras basically describe the Planck-scale-modified symmetry properties of particles moving, respectively, on a Minkowski or de Sitter spacetime. So it should come with no surprise that the same sort of effects emerges (appropriately modified) also in the Planck-scale-modified FLRW model we study here.

\subsubsection{Redshift}\label{ssec:redshift}
The energy redshift of a massless particle propagating in a qFLRW spacetime can be computed by studying the evolution of $p_t$ along the particle's worldline\footnote{Both for the computation of redshift in this subsection and of lateshift in subsection \ref{ssec:lateshift}, we do not need to worry about coordinates artifacts such as the ones discussed in \cite{Amelino-Camelia:2013uya}, because the spacetime coordinates we use are dual to the momenta from the point of view of Poisson brackets. If one were to choose a different spacetime coordinatization (such as, for example, spacetime coordinates inducing translations on momentum space coordinates), then one would need a more careful analysis, based on the techniques developed in the relative locality framework \cite{Amelino-Camelia:2013uya,Barcaroli:2015eqe}.}, eq. \eqref{eq:pt(t)massless}.

Note that in the classical limit $\ell \to 0 $ from \eqref{eq:pt(t)massless} one recovers the usual  scaling of the energy of a particle traveling in a FLRW universe: 
\begin{align}
	p_t(t)|_{\ell=0} = - \frac{w}{a(t)}\,.
\end{align}
The redshift of a massless particle on a radial geodesic is measured by comparing the energy at the time of emission, $p_t(t_i)$, with the energy at detection, $p_t(t_f)$:
\begin{align}\label{eq:redshift}
	z(t_i, t_f) &\equiv \frac{p_t({t_i})-p_t(t_f)}{p_t(t_f)} \nonumber\\ &=- \frac{\ell p_t({t_i})}{\ln\left(1- \frac{a(t_i)}{a(t_f)}(1-e^{-\ell p_t(t_i)})\right)} - 1\\ &=  \left(\frac{a(t_f)}{a(t_i)}-1\right)\left(1+\frac{\ell}{2} p_t(t_i)\right) + \mathcal{O}(\ell^2)\,.\nonumber
\end{align}
where we have used the fact that from eq. \eqref{eq:pt(t)massless} one can find the constant of motion $w$ as a function of the energy and scale factor at any given time \footnote{It is easy to check that from eq. \eqref{eq:w} one can indeed obtain $\frac{dw}{dt}=0$, consistently with $w$ being a constant of motion.}:
\begin{align}
w=- a(t)\frac{1-e^{-\ell p_t(t)}}{\ell}\,, \label{eq:w}
\end{align}
and in particular one can set $w=-a(t_i)\frac{1-e^{-\ell p_t(t_i)}}{\ell}$, so that
\begin{equation}
p_t(t)=-\frac{1}{\ell}\ln\left(1- \frac{a(t_i)}{a(t)}(1-e^{-\ell p_t(t_i)})\right) \overset{\ell\to 0 }{\longrightarrow} p_t(t_i) \frac{a(t_i)}{a(t)}\,. \label{eq:redshiftt}
\end{equation}
If the scale factor is the one of a de Sitter spacetime, $a(t) = e^{h t}$, where $h$ is the inverse of the de Sitter radius, the redshift is:
\begin{align}
	z(t_i, t_f) &=- \frac{\ell p_t({t_i})}{\ln\left(1- e^{h(t_i-t_f)}(1-e^{-\ell p_t(t_i)})\right)} - 1=  \left(e^{h(t_f-t_i)}-1\right)\left(1+\frac{\ell}{2} p_t(t_i)\right) +  \mathcal{O}(\ell^2)\,,
\end{align}
which reproduces the results found in \cite{Barcaroli:2015eqe} at the first order in $h$ and $\ell$ for a particle with $q$-de Sitter symmetries. So our result \eqref{eq:redshift} can be seen as the generalization of the formula for the redshift from a $q$-de Sitter model to a  $q$FLRW model. 

\subsubsection{Spacetime worldlines}\label{ssec:worldlines}
The worldline of a massless particle is found by integrating the equation for the evolution of the particle's radial coordinate, eq. \eqref{eq:rprimemassless}, which we rewrite here for convenience:
\begin{align}
r'(t)&= \frac{\chi a^{-1}}{\ell w a^{-1}+1}= \frac{\sqrt{1- k r^2}}{\ell w + a(t)}\,.
\end{align}
The constant $w$ depends on the particle energy and on the scale factor as in eq. \eqref{eq:w}. In particular, we can again choose to evaluate it at the initial time $t_{i}$, $w=- a(t_{i})\frac{1-e^{-\ell p_{t}(t_{i})}}{\ell}$. We also ask that the particle moves forward with time ($r$ increases with time).
The worldline can then be expressed as a formal integral:
\begin{align}
r(t)&=\frac{1}{\sqrt{k}}\sin\left( \arcsin\left(\sqrt{k}\; r(t_i)\right) + \sqrt{k}\int_{t_i}^{t} \frac{1}{\ell w + a(\tau)} d\tau   \right)\,.\label{eq:r(t)}
\end{align}
This can be evaluated explicitly once a  choice for the scale factor is made. For example if, as done in the end of the previous subsection, we choose the scale factor of an exponentially expanding universe, $a(t)=e^{h t}$, we find
\begin{align}\label{eq:r(t)k}
	r(t)
	&= \frac{1}{\sqrt{k}}\sin\left( \arcsin\left(\sqrt{k} r(t_i)\right) - \sqrt{k}\frac{e^{\ell p_t(t_i)-h t_i}}{h( e^{\ell p_t(t_i)} - 1)} \left[ h (t-t_i) - \ln\left( (e^{h(t-t_i)}- 1)e^{\ell p_t(t_i)}+1\right)\right] \right)\\
	&=\frac{1}{\sqrt{k}}\sin\left( \arcsin\left(\sqrt{k} r(t_i)\right) + \sqrt{k}( t - t_i)\right)\left( 1 + \frac{h}{2} k \ell p_t(t_i) (t + t_i)(t - t_i)^2\right)\nonumber \\
	&-\frac{( t - t_i) }{2}\cos\left( \arcsin\left(\sqrt{k} r(t_i)\right) + \sqrt{k}( t - t_i)\right) \big( h ( t + t_i ) + 2 \ell p_t(t_i) ( h t -1) \big) +\mathcal O(\ell^{2}, h^{2})\,.
\end{align}
The spatially-flat case is given by the $k=0$ limit of the above, and gives the Planck-scale-corrected worldline of a massless particle in de Sitter spacetime:
\begin{align}
\left[r(t)-r(t_i)\right]_{{dS, flat}}
&= \frac{e^{\ell p_t(t_i)-h t_i}}{h\left( 1-e^{\ell p_t(t_i)} \right)}\left[ h(t-t_i)-\ln\left( (e^{h(t-t_i)}- 1)e^{\ell p_t(t_i)}+1\right)   \right]\label{eq:r(t)flatdS}\\ 
&=  (t-t_{i}) \left(  1+\ell p_{t}(t_{i})-\frac{h}{2}\Bigg[t_{i}+t \Big(1+2 \ell p_{t}(t_{i})\Big)\Bigg]  \right) +\mathcal O(\ell^{2}, h^{2})\,.\label{eq:r(t)leadingorder}
\end{align}
Note that the first order in $h$ and $\ell$, reported in the second line, reproduces the results of \cite{Barcaroli:2015eqe}, which studied particles with symmetries described by the $q$-de Sitter Hopf algebra.

Already from the general expression \eqref{eq:r(t)} one can see that the distance travelled by a particle depends on its initial energy via the $w$ factor inside the integral. This energy-dependence starts at the first order in $\ell$, so it can be understood as a purely Planck-scale effect. The fact that in the $\ell, h$ expansion of eq. \eqref{eq:r(t)leadingorder} there are terms proportional to the product $h \ell$ indicates that the time of travel depends in a non-trivial way on the interplay between curvature and Planck-scale properties of spacetime (see also \cite{Marciano:2010gq}).

\subsubsection{Lateshift}\label{ssec:lateshift}
For phenomeological purposes, it is useful to compute the difference in arrival time of two massless particles which are emitted at the same time with different energies and are detected at the same location. This is in fact the most promising kind of observable that current astrophysical experiments are able to measure \cite{AmelinoCamelia:1997gz,Ellis:2009yx,Ackermann:2009aa,AmelinoCamelia:2009pg, Amelino-Camelia:2013naa, Amelino-Camelia:2015nqa}. One (soft) particle is assumed to have small momentum (such that $\ell$-dependent effects are negligible), while the other (hard) particle  is assumed to have high momentum (such that $\ell$-dependent effects are relevant). 

We compute the time delay at the first order in $\ell$ in two cases. The first one is a universe that is expanding exponentially at constant rate, $a(t)=e^{h t}$ with arbitrary spatial curvature $k$. The second option we consider is a universe with arbitrary scale factor $a(t)$ and vanishing spatial curvature,~$k=0$.

In the first case the worldlines of massless particles are given by eq. \eqref{eq:r(t)k}. To compute the time delay we set $t_i=0$, $r(t_i)=0$ in eq. \eqref{eq:r(t)k} and equate the distances covered by the two particles, respectively during the time $t^{hard}$ and $t^{soft}$:
\begin{align}\label{eq:rh=rs}
	 r^{hard}(t^{hard})=r^{soft}(t^{soft}) \,.
\end{align} 
These distances are given by the following expressions up to first order in $\ell$
\begin{align}
	r^{hard}(t^{hard})&=\frac{1}{\sqrt{k}}\sin \left(\sqrt{k}\; \frac{1-e^{-h t^{hard}}}{h}\right) + \ell\; \frac{p_t(0)}{2 h}\left(1-e^{-2 h t^{hard}}\right) \cos\left(\sqrt{k}\frac{1-e^{-h t^{hard}}}{h}\right)  \,,\\
	r^{soft}(t^{soft})&=\frac{1}{\sqrt{k}}\sin \left(\sqrt{k}\; \frac{1-e^{-h t^{soft}}}{h}\right)\,,
\end{align}
which realize the idea that the soft photon does not feel any $\ell$-dependent deformation. Then the difference in time of arrival, $\Delta t\equiv t^{hard}-t^{soft}$ can be calculated to first order in $\ell$\footnote{Simply replace $t^{soft} = t^{hard} - \ell \delta t$ in \eqref{eq:rh=rs}, solve for $\delta t$ to zeroth order in $\ell$ and identify $\Delta t = \ell \delta t$.} and can be expressed in terms of the redshift of the hard particle, $z(0, t^{hard})=e^{ h t^{hard}} - 1+\mathcal{O}(\ell)$ \footnote{We approximate the redshift at zeroth order in $\ell$ since it will appear in terms which are already first order in $\ell$. }, and its energy at the detector, $p_t(t^{hard})=p_t(0)/(1+z)$, as follows:
\begin{align}\label{eq:DT}
	\Delta t|_{a(t)=e^{h t}}&= - \frac{\ell}{h} p_t(t^{hard})\left(z+\frac{z^2}{2}\right) +\mathcal O(\ell^2)\,.
\end{align}
Surprisingly, we find that $\Delta t$ is independent of the spatial curvature $k$ up to first order in $\ell$. 

In the second case, the relevant expression of the worldline is given by eq. \eqref{eq:r(t)}, with $k=0$. Then the distance travelled by the two particles, respectively during the time $t^{hard}$ and $t^{soft}$, is\footnote{We write the constant $w$ as a function of energy and scale factor measured at the detection time of the hard particle: $w=a(t^{hard}) \frac{1-e^{-\ell p_t(t^{hard})}}{\ell}$.}
\begin{align}
	r^{hard}(t^{hard})&=\int_{0}^{t^{hard}} \frac{d\tau}{ a(\tau)}+\ell p_t(t^{hard}) \,a(t^{hard})\int_{0}^{t^{hard}} \frac{d\tau}{ a(\tau)^2}\,,\\
	r^{soft}(t^{soft})&=\int_{0}^{t^{soft}} \frac{d\tau}{ a(\tau)}\,.
\end{align}
Following the same procedure as above, we equate these distances, obtaining
\begin{align}
	\int_{t^{hard}}^{t^{soft}}\frac{d\tau}{ a(\tau)}&= \ell p_t(t^{hard})\, a(t^{hard})\int_0^{t^{hard}}\frac{d\tau}{ a(\tau)^2}\,.
\end{align}
Since the variation of $a(t)$ is negligible between $t^{hard}$ and $t^{soft}$, we can solve this  equality and write the time delay as:
\begin{align}\label{eq:DTgeneric}
	\Delta t|_{a(t),k=0}&= -\ell p_t(t^{hard})\int_0^{t^{hard}}\frac{d\tau}{ a(\tau)^2}\,.
\end{align}

These two results on the time delay can be compared with the ones of ref. \cite{Rosati:2015pga}. There, the time delay was computed both assuming an exponential scale factor with vanishing spatial curvature ($a(t)=e^{h t}$, $k=0$) and for a generic scale factor and vanishing spatial curvature. For both of the choiches of scale factors, the Planck-scale corrections were derived from two different models: one that explicitly breaks Lorentz symmetries (LIV case) and another one where the Planck-scale corrections are compatible with deformed relativistic symmetries (DSR case).

In the model with exponential scale factor the time delay is given in eqs. (18) and (30) of \cite{Rosati:2015pga}, respectively for the  LIV and DSR cases. 
We rewrite these equations here for convenience, adapting the notation to ours: 
\begin{align}
	\Delta t|_{a(t)=e^{h t},k=0}^{LIV}&=\frac{p^{hard}}{h}\left( \lambda'\ln(1+z)+\lambda''z+\lambda (z+\frac{z^2}{2})+\lambda'''(z+z^2+\frac{z^3}{3})  \right)\,,\\
	\Delta t|_{a(t)=e^{h t},k=0}^{DSR}&=\frac{\ell \,p^{hard}}{h}\left( (\alpha-\gamma)\ln(1+z)+(\beta+\gamma)(z+\frac{z^2}{2})  \right)\,.
\end{align}
These equations contain a number of parameters, which can be chosen so as to match our result, eq. \eqref{eq:DT}. In particular, for the LIV case this choice of parameters is $\{\lambda'=\lambda''=\lambda'''=0,\,\lambda=-\ell\}$, while for the DSR case one has to fix $\{\alpha=\gamma,\,\beta+\gamma=-1\}$.

In the model with generic  scale factor the time delay is obtained in ref. \cite{Rosati:2015pga} via a  procedure that describes FLRW spacetimes as the "glueing" of many slices of de Sitter spacetime with different values of $h$. The resulting time delay is given in eqs. (65) and the one before (81) of \cite{Rosati:2015pga}, again respectively for the  LIV and DSR cases. We rewrite here also these equations, adapting the notation to ours:
\begin{align}
	\Delta t|_{a(t),k=0}^{LIV}&=  p^{hard}\int_0^{t^{hard}}d\tau\left[\lambda'+\frac{\lambda''}{a(\tau)}+\frac{\lambda}{a(\tau)^2}+\frac{\lambda'''}{a(\tau)^3}\right] \,,\\
	\Delta t|_{a(t),k=0}^{DSR}&= \ell p^{hard}\left((\beta+\gamma)\int_0^{t^{hard}}\frac{d\tau}{ a(\tau)^2}+(\alpha-\gamma)\int_0^{t^{hard}}d\tau\left(a(\tau)^{-1}-\frac{a'(\tau)}{a(\tau)}\int_{\tau}^{t^{hard}}\frac{d\tau'}{a(\tau')}\right)\right)\,.
\end{align}
Of course these equations depend on the same parameters as the equations for the exponential scale factor case, since they result from the glueing procedure described above. Our expression for the time delay in a universe with generic scale factor, eq. \eqref{eq:DTgeneric}, matches the results of \cite{Rosati:2015pga} for the same choice of parameters as in the de Sitter case.
This is a confirmation of the actual accuracy of the "glueing" procedure used in \cite{Rosati:2015pga}, since our computation does not rely on any approximation of this sort.

\section{Discussion}
Hamilton geometry allows to consistently introduce departures from general-relativistic geodesic motion as a consequence of the presence of non-quadratic functions of the  four momentum of a test particle in its Hamiltonian.
While in this framework the Hamiltonian is, in principle, allowed an arbitrary dependence on momentum, symmetry requirements constrain its possible shape. In the work presented here, the implementation of spatial isotropy and homogeneity as invariances of the Hamiltonian allowed us to derive its generic functional form in the cosmological setting, eq. (\ref{eq:generalH}).

The presence of terms which are of higher power in the momenta requires the introduction of a dimensionful  parameter, i.e. of an energy scale, so this framework is a promising candidate to describe Planck-scale departures from general relativity. In particular, since Hamilton geometry allows for a consistent treatment of the symmetry properties of the modified phase space, it can be used for the scopes of the 'Doubly Special Relativity' program.
Presently, the most successful theoretical framework for relativistic Planck-scale-modified dispersion relations is the one of Hopf algebras \cite{Majid:1994cy}, which is however rooted in the assumption of maximal symmetry.
The strength of our approach lies in its ability to deal with situations of only partial symmetry, as it is the case in cosmology, which is extremely relevant for Planck-scale phenomenology.

We provided the most general Hamiltonian that modifies the one valid in a Friedmann-Lema\^itre-Robertson-Walker universe, up to the first order in the (dimensionful) deformation parameter, meaning that our Hamiltonian contains up to cubic powers in the four-momentum. This is displayed in eq. \eqref{eq:dFLRW}.

Most importantly we were able to propose a Hamiltonian which is a curved-spacetime generalization, compatible with spacetime isotropy and homogeneity, of the $\kappa$-Poincar\'e dispersion relation - one of the most studied deformations of the special-relativistic dispersion relation, which is derived in the context of the $\kappa$-Poincar\'e Hopf algebra. Our proposal, eq. \eqref{eq:qFLRW}, is compatible with the loss of time-translation invariance in the cosmological setting and reduces to the $\kappa$-Poincar\'e one when going to a local inertial frame. This was made especially easy by the fact that the $\kappa$-Poincar\'{e} Casimir exhibits the same spatio-temporal split as results from cosmological symmetry.

Finally, we used this last Hamiltonian to study in detail the worldlines of relativistic particles and compute some observable predictions.
Specifically, we were able to reproduce and generalize the familiar energy redshift effect, eq. \eqref{eq:redshift},  computed to all orders in the Planck length and for arbitrary scale factor and  spatial curvature. In general relativity such effect only depends on the time of travel of the particle, while in our framework it also depends on the particle's energy itself. Choosing the scale factor of an exponentially expanding universe and considering only the first order perturbation in the Planck length, our result confirms the results found in \cite{Barcaroli:2015eqe}. 
We were also able to compute, up to the first order in the Planck length, the amount of delay in the time of arrival of two particles emitted by the same source at the same time but with different energies,~eq.~\eqref{eq:DTgeneric}.  This is an effect which is being currently tested with astrophysical observations \cite{Ackermann:2009aa, Amelino-Camelia:2013naa, Amelino-Camelia:2015nqa, Xu:2016zsa, Amelino-Camelia:2016fuh}.
We could confirm the results of \cite{Rosati:2015pga}, valid for a spatially flat universe, for a specific choice of the parameters in \cite{Rosati:2015pga}. Our results also generalise the computation to a universe with arbitrary spatial curvature.

In an upcoming article we will use the Hamilton geometry framework to study  modified dispersion relations in  arbitrary curved spacetimes. In the same article we will then introduce spherically symmetric Planck-scale-modified geometries, which can be interpreted as deformations of Schwarzschild geometry.

\begin{acknowledgments}
GG acknowledges support form the John Templeton Foundation. NL acknowledges support by the European Union Seventh Framework Programme (FP7 2007-2013) under grant agreement 291823 Marie Curie FP7-PEOPLE-2011-COFUND (The new International Fellowship Mobility Programme for Experienced Researchers in Croatia - NEWFELPRO), and also partial support by the H2020 Twinning project n$^\text{o}$ 692194, RBI-TWINNING and by the 000008 15 RS {\it Avvio alla ricerca} 2015 fellowship (by the italian ministry of university and research). LKB is supported by a Ph.D. grant of the German Research Foundation within its Research Training Group 1620, \emph{Models of Gravity}. CP gratefully thanks the Center of Applied Space Technology and Microgravity (ZARM) at the University of Bremen for their kind hospitality. LKB and CP would also like to thank their colleague Michael Fennen for critical remarks and important discussions.
\end{acknowledgments}

\newpage
\appendix
\newpage

\section{Complete lifts of the symmetry generators}\label{app:CL}
In section \ref{sec:homiso} the symmetry conditions on dispersion relations are formulated in terms of complete lifts of the symmetry vector fields \eqref{eq:KVF1}-\eqref{eq:KVF2} on spacetime. Here we display these lifts explicitly.

Defining $\chi = \sqrt{1-kr^2}$ the lifted translations are:
\begin{align}
	X_1^C &= \chi \sin\theta \cos\phi \partial_r + \frac{\chi}{r} \cos\theta \cos\phi \partial_\theta - \frac{\chi}{r}\frac{\sin\phi}{\sin\theta}\partial_\phi \nonumber\\
			   &+ \bigg(\frac{kr}{\chi} \sin\theta \cos\phi p_r + \frac{1}{\chi r^2} \cos\theta \cos\phi p_\theta - \frac{1}{\chi r^2}\frac{\sin\phi}{\sin\theta} p_\phi\bigg)\bar{\partial}^r\\
			   &+ \bigg(- \chi \cos\theta \cos\phi p_r + \frac{\chi}{r} \sin\theta \cos\phi p_\theta - \frac{\chi}{r}\frac{\cos\theta}{\sin\theta^2} \sin\phi p_\phi\bigg)\bar{\partial}^\theta\nonumber\\
			   &+ \bigg( \chi \sin\theta \sin\phi p_r + \frac{\chi}{r} \cos\theta \sin\phi p_\theta + \frac{\chi}{r}\frac{\cos\phi}{\sin\theta} p_\theta\bigg)\bar{\partial}^\theta\,,\nonumber
\end{align}
\begin{align}
	X_2^C &= \chi \sin\theta \sin\phi \partial_r + \frac{\chi}{r} \cos\theta \sin\phi \partial_\theta + \frac{\chi}{r}\frac{\cos\phi}{\sin\theta}\partial_\phi\nonumber\\
	&+ \bigg(\frac{kr}{\chi} \sin\theta \sin\phi p_r + \frac{1}{\chi r^2} \cos\theta \sin\phi p_\theta + \frac{1}{\chi r^2}\frac{\cos\phi}{\sin\theta} p_\phi\bigg)\bar{\partial}^r\\
	&+ \bigg(- \chi \cos\theta \sin\phi p_r + \frac{\chi}{r} \sin\theta \sin\phi p_\theta + \frac{\chi}{r}\frac{\cos\theta}{\sin\theta^2} \cos\phi p_\phi\bigg)\bar{\partial}^\theta\nonumber\\
	&+ \bigg( -\chi \sin\theta \cos\phi p_r - \frac{\chi}{r} \cos\theta \cos\phi p_\theta + \frac{\chi}{r}\frac{\sin\phi}{\sin\theta} p_\theta\bigg)\bar{\partial}^\theta\,,\nonumber
\end{align}
\begin{align}
	X_3^C &= \chi \cos\theta \partial_r - \frac{\chi}{r}\sin\theta \partial_\theta\nonumber\\
			   &+ \bigg(\frac{kr}{\chi} \cos\theta p_r - \frac{1}{\chi r^2}\sin\theta p_\theta\bigg)\bar{\partial}^r\\
			   &+ \bigg(\chi \sin\theta p_r + \frac{\chi}{r} \cos\theta p_\theta \bigg)\bar{\partial}^\theta\,,\nonumber
\end{align}
while the lifted rotations are
\begin{align}
	X_4^C &= \sin\phi \partial_\theta + \cot\theta \cos\phi \partial_\phi\nonumber\\
			   &+ \frac{\cos\phi}{\sin\theta^2}p_\phi \bar{\partial}^\theta - \bigg(\cos\phi p_\theta - \cot\theta \sin\phi p_\phi\bigg)\bar{\partial}^\phi\,,
\end{align}
\begin{align}
	X_5^C &= -\cos\phi \partial_\theta + \cot\theta \sin\phi \partial_\phi\nonumber\\
	&+ \frac{\sin\phi}{\sin\theta^2}p_\phi \bar{\partial}^\theta - \bigg(\sin\phi p_\theta + \cot\theta \cos\phi p_\phi\bigg)\bar{\partial}^\phi\,,
\end{align}
\begin{align}
	X_6^C &= \partial_\phi\,.
\end{align}

\section{Solving the symmetry conditions}\label{app:symcon}
As mentioned in section \ref{sec:homiso} we can solve the symmetry conditions $X^C_I(H)=0$ for $I=1, ..., 6$ as partial differential equations by the introduction of coordinates
\begin{align}
	\hat t = t,\ \hat r = r,&\ \hat \theta = \theta,\ \hat \phi = \phi,\ \hat p_t = p_t,\ \hat p_r = p_r,\ \hat p_\theta = p_\theta,\\
	w^2 &= (1 - kr^2)p_r^2 + \frac{1}{r^2}p_\theta^2 + \frac{1}{r^2\sin\theta^2}p_\phi^2\,.
\end{align}
We replace all the $p_\phi$ in the expressions of the $X^C_I$, which are displayed in appendix~\ref{app:CL}, with
\begin{align}
	p_\phi^2 = r^2 \sin\theta^2 \bigg(w^2 - (1-kr^2)p_r^2 - \frac{1}{r^2}p^2_\theta\bigg)\,,
\end{align}
and we transform the coordinate basis of the vectors fields according to
\begin{align}
	\partial_t &= \hat \partial_t\,,\ \bar{\partial}^t = \hat{\bar{\partial}}^t\,, \\
	\partial_r &= \hat \partial_r - \bigg(\frac{w^2}{r} - \bigg(\frac{1}{r}-2kr\bigg)p_r^2\bigg)\frac{1}{w}\partial_w\,,\\
	\bar{\partial}^r &= \hat {\bar\partial}^r + (1-kr^2)p_r \frac{1}{w}\partial_w\,,\\
	\partial_\theta &= \hat \partial_\theta - \cot\theta (w^2 - (1-kr^2)p_r^2 - \frac{1}{r^2}p_\theta^2)\frac{1}{w}\partial_w\,,\\
	\bar{\partial}^\theta &= \hat {\bar\partial}^\theta + \frac{p_\theta}{r^2}\frac{1}{w}\partial_w\,,\\
	\partial_\phi &= \hat \partial_\phi\\
	\bar{\partial}^\phi &= \frac{1}{w}\frac{1}{r \sin\theta}\sqrt{w^2 - (1-kr^2)p_r^2-\frac{1}{r^2}p_\theta^2} \partial_w\,.
\end{align}
Expanding the symmetry equations $X^C_I(H)=0$ in these new coordinates we find
\begin{align}
	\hat \partial_r H = 0,\ \hat \partial_\theta H = 0,\ \hat \partial_\phi H = 0,\ 
	\hat {\bar{\partial}}^r H = 0,\ \hat {\bar{\partial}}^\theta H = 0\,.
\end{align}
The calculation is straightforward however quite lengthy. One trick is to consider the equations $(\sin\phi X^C_5 + \cos\phi X^C_4)H=0$,  $(\cos\phi X^C_5 - \sin\phi X^C_4)H=0$, $(\sin\phi X^C_1 - \cos\phi X^C_2)H=0$ as well as $(\cos\phi X^C_1 + \sin\phi X^C_2)H=0$ instead of $X^C_5 H=0$, $X^C_4 H=0$, $X^C_2 H=0$ and $X^C_1 H=0$. 

\section{Extrinsic viewpoint on homogeneity and isotropy conditions}\label{sec:App-ExtrView}
What underpins the postulate of isotropy and spatial homogeneity from a  
group-theoretical point of view is particularly visible when adopting the following extrinsic viewpoint.

The assumption of maximal spatial symmetry means that spacetime is foliated in 
time by three-dimensional, homogeneous submanifolds $\Sigma$, which are
necessarily isometric to either of the following:
\begin{align}
	\Sigma \cong 
	\begin{cases}
		\mathrm S^3 \cong \mathrm{SO(4)/SO(3)} 
				\quad \text{(a sphere)  if}\quad k=+1, \\
		\mathrm E^3 \cong \mathrm{ISO(3)/SO(3)}
				\quad \text{(Euclidean space)  if}\quad k=0, \\
		\mathrm H^3	 \cong \mathrm{SO(1,3)/SO(3)}
				\quad \text{(hyperbolic space)  if}\quad k=-1.
	\end{cases}
\end{align}
Here, the parameter $k$ indicates spatial curvature, and the identification as 
either of the three given quotient spaces is understood from an embedding of
$\Sigma$ into $\mathbb R^4$ as
\begin{align}\label{eq:embed}
	G_{MN}X^M X^N = k^{-1}
\end{align}
with metric $G=\mathrm{diag}(1,1,1,k^{-1})$, in Cartesian coordinates 
$X^M$ (indices $M,N,\dots$ between 1 and 4), where the case $k=0$ is  
considered only in the limiting sense. In the case $\Sigma\cong \mathrm S^3$, 
choosing
\begin{align}
	X^M = 	\begin{pmatrix} 
				\sin \rho \sin \theta \cos \phi \\
				\sin \rho \sin \theta \sin \phi \\
				\sin \rho \cos \theta \\
				\cos \rho
			\end{pmatrix}\quad \text{for}\quad k=+1,
\end{align}
induces on $\Sigma$ the metric
\begin{align}
	h = \mathrm d \rho^2 + \sin^2\!\rho\; \mathrm d \Omega_2^2
    = (1-r^2)\mathrm d r^2 + r^2\mathrm d \Omega_2^2
\end{align}
with $\mathrm d \Omega_2^2 = \mathrm d\theta^2 + \sin^2\!\phi$, and, 
optionally, $r=\sin \rho$.
The remaining two cases are completely analogous. (Formally, $k=-1$ is obtained 
by sending $\rho\rightarrow i\rho$, and $k=0$ when neglecting all but linear 
terms in $\rho$).
The isometries are generated by the restriction of the vector fields 
$Z_{ij}$ and $k Z_{4i}$ ($i,j=1,2,3$) to $\Sigma$, where
\begin{align} \label{eq:EmbeddedKilling}
	Z_{MN} = 2 X^L G_{L[M} \partial_{N]}.
\end{align}
Cotangent vectors on $\Sigma$ have, in this embedded view, components
\begin{align}\label{eq:EmbeddedCovecComp}
	P_M = \frac{\partial \rho}{\partial X^M}p_\rho + \frac{\partial 
	\theta}{\partial X^M}p_\theta + \frac{\partial \phi}{\partial X^M}p_\phi.
\end{align}
Lifting the vector fields \eqref{eq:EmbeddedKilling} to $T^*\mathbb R^4 \cong 
\mathbb R^8$ simply yields
\begin{align}
	\hat Z_{MN} = Z_{MN} + 2 P_{[M}G_{N]L}\bar\partial^L,
\end{align}
the action of which can be represented as
\begin{align}\label{eq:action}
	\delta_{\hat Z_{MN} } Y = \Omega_{MN} Y
\end{align}
with
\begin{align}
	\Omega_{MN} &=     	\begin{pmatrix}
						e_N e^*_M - e_M e^*_N & 0 \\
            			0& (e_M e^*_N - e_N e^*_M)^t
						\end{pmatrix} \\
	\text{and}\qquad
	Y &= (X^1,\dots,X^4,P_1,\dots,P_4)^t,
\end{align}
when $e_M$ is the $\mathbb R^4$ unit (column) vector in the $M$-th direction, 
and $e^*_M =G(e_M)$ the dual (row) vector. Since both the lift and the matrix representation are linear operations, the rescaling $Z_{4i}\rightarrow k Z_{4i}$ translates identically to $\Omega_{4i}\rightarrow k \Omega_{4i}$ to yield the correct $k\rightarrow 0$ limit. Respecting this, any function invariant under \eqref{eq:action} must be a function of some
quadratic form $Y^t h Y$, with symmetric coefficient matrix $h$. The condition it 
has to satisfy is 
\begin{align}
	\Omega_{MN}^t h + h\Omega_{MN}^{~} = 0,
\end{align}
which is the infinitesimal version of $\Omega_{MN}^t h \Omega_{MN}=h$. It is
solved if
\begin{align}
	h\in \mathrm{span}\Bigg\{ 
	\begin{pmatrix}
		kG & 0 \\ 0 & 0
	\end{pmatrix},
	\begin{pmatrix}
		0 & 0 \\ 0 & G^{-1}
	\end{pmatrix},    
	\begin{pmatrix}
		0 & \mathbb 1 \\ \mathbb 1 & 0
	\end{pmatrix}
	\Bigg \},
\end{align}
Now, $X^M P_M=0$ on $\Sigma$, and $G_{MN}X^M X^N=k^{-1}=\mathrm{const.}$ anyway,
so that a generic function invariant under the symmetry-generating vector fields
$\hat Z_{MN}$ can only truly depend on $(G^{-1})^{MN}P_M P_N$. But on $T^*\Sigma$, 
one easily finds, with $w$ from \eqref{eq:w2def}, that
\begin{align}
w^2 = (G^{-1})^{MN}P_M P_N.
\end{align}
This is the aimed-at result. Considering a time-dependent embedding \eqref{eq:embed} 
then creates the additional dependence on the corresponding coordinates
$t$ and $p_t$ for a general homogenous and isotropic Hamiltonian \eqref{eq:generalH}. 


\section{More on the form of the qFLRW Hamiltonian}
The {\it ansatz} we formulated at the beginning of section \ref{sec:qFLRW}, for what concerns the form of the Hamiltonian, can be derived formally from the five-dimensional $\kappa$-Poincar\'{e} momentum-space \cite{Sitarz:1994rh,Agostini:2002yd}. It is well argumented in the literature that $\kappa$-Poincar\'{e} (bicrossproduct basis) Hopf algebra features can be reproduced as a curved momentum-space Riemannian geometry with a de Sitter-like metric \cite{AmelinoCamelia:2011nt,Gubitosi:2013rna,Amelino-Camelia:2013sba,Loret:2014prd}. Such a curved momentum-space (here discussed in 1+1D for the sake of simplicity) can be realized as a 1+1D hyperboloid embedded into a 1+1+1D Minkowskian manifold with coordinates $\Pi_0,\, \Pi_1, \Pi_4$, which on the hyperboloid satisfy the constraint 
\begin{equation}
-\Pi_0^2 + \Pi_1^2 + \tilde{\Pi}_4^2 = \frac{1}{\ell^2}\,.
\end{equation}
We can redefine the coordinate $\tilde{\Pi}_4 -1/\ell=\Pi_4 $, in order to reabsorb the term $1/\ell^2$.
This set of Minkowskian 1+1+1D spacetime coordinates are related to the 1+1D corrdinates $p_\mu$ defined on the hyperboloid by the following relations:
\begin{eqnarray}
\Pi_0&=&\frac{1}{\ell}\sinh(\ell p_t)+\frac{\ell}{2}e^{\ell p_t}p_1^2\,,\nonumber\\
\Pi_1&=& p_1 e^{\ell p_t}\,,\\
\Pi_4&=&\frac{1}{\ell}(\cosh(\ell p_t)-1)-\frac{\ell}{2}e^{\ell p_t}p_1^2\,.\nonumber
\end{eqnarray}
For $\eta^{AB}$ being the 1+1+1D Minkowskian metric with signature $(-1,1,1)$, we have
\begin{equation}
\mathcal{H}_\kappa=\eta^{AB}\Pi_A\Pi_B=-\Pi_0^2+\Pi_1^2+\Pi_4^2=
-\frac{4}{\ell^2}\sinh^2\bigg(\frac{\ell}{2} p_t\bigg) + e^{\ell p_t}\delta^{\alpha\beta}\tilde p_\alpha \tilde p_\beta \,.\label{eq:5DCasimir}
\end{equation}
The physical interpretation of such momentum-space geodesic equations can be formalized interpreting the Hamiltonian as a "momentum-space invariant line-element":
\begin{equation}
\mathcal{H}_\kappa=\int_0^1\eta^{AB}\dot{\Pi}_A(s)\dot{\Pi}_B(s)\, ds=\int_0^1\zeta^{\alpha\beta}(P)\dot{P}_\alpha(s)\dot{P}_\beta(s)\, ds\, ,\label{eq:kCasimirFromMetric}
\end{equation}
where the $P_\alpha (s)$ are momentum-space geodesics and $\zeta^{\mu\nu}$ is the 1+1D momentum-space metric, defined as
\begin{equation}
\zeta^{\alpha\beta}(p)=\eta^{AB}\frac{\partial \Pi_A}{\partial p_\alpha}\frac{\partial \Pi_B}{\partial p_\beta}=
\left(\begin{array}{cc}
-1 & 0\\
0 & e^{2\ell p_t}
\end{array}\right)\,.\label{eq:momspmetric}
\end{equation}

This simple observation allows us to define a  {\it deSitter-deSitter} framework in which Hamiltonian and metric can be obtained from (\ref{eq:5DCasimir}) and (\ref{eq:momspmetric}), just sending $p_1\rightarrow p_1 e^{-h t}$, {\it id est}
\begin{eqnarray}
\Pi^{dS-dS}_0&=&\frac{1}{\ell}\sinh(\ell p_t)+\frac{\ell}{2}e^{\ell p_t}e^{-2 h t}p_1^2\,,\nonumber\\
\Pi^{dS-dS}_1&=& p_1 e^{\ell p_t-h t}\,,\\
\Pi^{dS-dS}_4&=&\frac{1}{\ell}(\cosh(\ell p_t)-1)-\frac{\ell}{2}e^{\ell p_t}e^{-2 h t}p_1^2\,.\nonumber
\end{eqnarray}
The Hamiltonian then becomes
\begin{equation}
{\cal H}_{dS-dS}=-\frac{4}{\ell^2}\sinh^2\left({\frac{\ell p_t}{2}}\right)+e^{\ell p_t}e^{-2 h x^0}p_1^2\,,\label{eq:dSdSCasimir}
\end{equation}
and the metric of momentum-space appears to be a deSitter-deSitter like one
\begin{equation}
\zeta_{dS-dS}^{\alpha\beta}(p)=\eta^{AB}\frac{\partial \Pi^{dS-dS}_A}{\partial p_\alpha}\frac{\partial \Pi^{dS-dS}_B}{\partial p_\beta}=
\left(\begin{array}{cc}
-1 & 0\\
0 & e^{-2 h x^0}e^{2\ell p_t}
\end{array}\right)\,.\label{eq:dSdSmommetric}
\end{equation}

The extra-dimension formalism can be very useful to deal with deSitter momentum-space geodesic equations 
\begin{equation}
\ddot{P}_\alpha+\Gamma^{\beta\gamma}_\alpha\dot{P}_\beta\dot{P}_\gamma=0\,,\label{eq:GeoEqmomsp}
\end{equation} 
where the $P_\mu(s)$ are the momentum-space geodesics in 1+1D and the dotted coordinate $\dot{P}$ is differentiated with respect a geodesic affine parameter $s$. Those equations are in general very complex to solve, as we will see in a few lines.
Using the metric (\ref{eq:momspmetric}), geodesic equations (\ref{eq:GeoEqmomsp}) become
\begin{equation}
\left\{\begin{array}{l}
\ddot{P}_0+\ell e^{2\ell P_0}\dot{P}_1^2=0\\
\ddot{P}_1+2\ell\dot{P}_0\dot{P}_1=0\,\label{eq:GeoEqmom}
\end{array}\right.
\end{equation}
with conditions $P_\alpha(0)=0$ and $P_\alpha(1)=p_\alpha$. Those latter equations (in which the connection is the usual Christoffel symbol in terms of the momentum-space metric), as previously observed, are not so easy to solve. However being the 1+1+1D metric flat, we can take advantage of $\ddot{\Pi}_\alpha=0$ and then $\dot{\Pi}_\alpha=const.$, which give us a couple of simple differential equations:
\begin{equation}
\left\{\begin{array}{l}
\dot{\Pi}_0+\dot{\Pi}_4=e^{\ell P_0}\dot{P}_0=C_{04}\\
\dot{\Pi}_1=\ell e^{\ell P_0}\dot{P}_0 P_1+e^{\ell P_0}\dot{P}_1=C_1
\end{array}\right.\,,
\end{equation}
whose solutions
\begin{equation}
\left\{\begin{array}{l}
P_0(s)=-\frac{1}{\ell}\ln\left(1+(e^{\ell p_t}-1)s\right)\\
P_1(s)=\frac{e^{\ell p_t} p_1 s}{1+(e^{\ell p_t}-1)s}
\end{array}\right.\,,
\end{equation}
always satisfy the first equation in (\ref{eq:GeoEqmom}) and also the second one, if we take into account the additional on-shell condition. A further check can be fulfilled by using those latter solutions to calculate the $\kappa$-Poincar\'{e} Hamiltonian with the procedure defined in eq.\eqref{eq:kCasimirFromMetric}.\\
The possibility to describe $\kappa$-Poincar\'{e} Hopf algebra formalism as a curved momentum-space allows us to establish a duality between complex Hopf algebra properties and Riemannian geometry, which makes it possible to interpret in a comprehensible way all the otherwise obscure $\kappa$-Poincar\'{e} features like the time-delay effect and nontrivial composition law for momenta \cite{Amelino-Camelia:2013uya}.

\bibliographystyle{utphys}
\bibliography{HI}

\end{document}